\documentclass[twocolumn,showpacs,preprintnumbers,amsmath,amssymb]{revtex4}
\usepackage{graphicx}% Include figure files
\usepackage{dcolumn}% Align table columns on decimal point
\usepackage{bm}% bold math

\newcommand {\be}{\begin{equation}}
\newcommand {\ee}{\end{equation}}
\newcommand {\bea}{\begin{eqnarray}}
\newcommand {\eea}{\end{eqnarray}}
\newcommand {\bem}{\begin{displaymath}}
\newcommand {\eem}{\end{displaymath}}
\newcommand {\f}{\frac}
\newcommand {\p}{\partial}
\begin{document}

\preprint{ }

\title{Emergence of dissipative structures in current-carrying superconducting wires.}% Force line breaks with \\
\author{ G. A. Levin, P. N. Barnes, J. P. Rodriguez*, J. A. Connors** and J. S. Bulmer }
%\altaffiliation[Also at ]{Physics Department, XYZ University.}%Lines break %automatically or can be forced with \\
%\author{ J. S. Bulmer, J. A. Connors }
\affiliation{Air Force Research Laboratory, Propulsion Directorate, Wright-Patterson Air Force Base, OH 45433. \\
*Department of Physics and Astronomy, California State University, Los Angeles, CA 90032\\
**Department of Physics, Ohio State University, Columbus, OH, USA; }
%
%\author{ J. P. Rodriguez }
%\affiliation{ Department of Physics and Astronomy, California State University, %Los Angeles, CA }

\date{\today}% It is always \today, today,
             % but any date may be explicitly specified

\begin{abstract}
We discuss the emergence of a spontaneous temperature and critical current spatial modulation in current-carrying high temperature superconducting wire. The modulation of the critical current along the wire on a scale of $3 - 10\; mm$ forces a fraction of the transport current to crisscross the resistive interface between the superconducting film and normal metal stabilizer attached to it. This generates additional heat that allows such a structure to be self sustainable. Stability and the conditions for experimental observation of this phenomenon are also discussed. \end{abstract}
\pacs{74.72.-h, 85.25.-j, 05.65.+b, 05.45.-a, 74.90.+n}
\maketitle

\section{\label{sec:level1}Introduction \protect}

Propagation of electric current through a superconducting substance is an inherently unstable process. If a small section of the superconductor turns normal, the heat generated by the current may lead to a thermal runaway process (quench). In practical (stabilized) superconducting wires a normal conductor (stabilizer) is electrically and thermally coupled to the superconducting material. This allows the wire to recover from accidental temperature perturbations or, if stabilization fails, to quench without irreparable damage\cite{Iwasa}. Thus, a current-carrying stabilized superconducting wire is a bistable system with two spatially uniform stable modes of operation: one is superconducting -- low temperature and low (practically zero) dissipation, the other is normal -- high temperature (greater than the critical temperature) and high dissipation. A transition from superconducting to normal mode initiated by a local temperature perturbation leads to a normal zone propagation (NZP) characterized by a certain speed. 

Here we will discuss the transition from superconducting to normal mode of operation in the state-of-the art $YBa_2Cu_3O_{7-x}$ ($YBCO$) coated conductors\cite{Larbalestier,Foltyn}.  The main advantage of $YBCO$ coated conductors over conventional low temperature superconducting wires -- high operating temperature ($65-77\;K$) -- has an undesirable flip side: their heat capacity is high in comparison with that of low-$T_c$ superconducting wires. As the result, the normal zone propagates slowly which significantly complicates protection of the devices, such as magnets or cables, from quench induced damage. We have investigated a possibility to increase the NZP speed by inserting a thin resistive layer (interface) between the superconducting film and the stabilizer. There is always a very thin (a fraction of a micron) layer of material between the $YBCO$ and copper that accounts for the resistance to the current exchange\cite{Ekin,Levin}. Conventional wisdom is that the interface resistance has to be as low as possible. If desired, as our findings suggest, it can be readily increased by various means. We have shown that the NZP speed can be substantially increased by increasing the interface resistance. This is the result of the greater amount of heat generated in the interface during current transfer from superconductor to the stabilizer\cite{ASC}. Similar conclusions have been drawn in a large body of work, mostly theory, devoted to the effects of interfacial resistance on NZP in conventional superconductors\cite{A,G,Gurevich}.  

The subject of this paper is a finding that when the interface resistance exceeds a certain threshold, the conventional scenario of transition between the superconducting and normal modes of operation breaks down and a current-carrying superconducting wire may exhibit a much more complex behavior than it is commonly expected. In addition to two uniform states -- superconducting and normal -- there is also an anomalous nonuniform, stable mode of operation characterized by a spontaneously developing pattern of spatial temperature modulation along the length of the wire with a peak temperature below the critical temperature $T_c$ of the superconductor. The temperature variation modulates the critical current density, forming the static temperature and critical current density ripples ($T$-ripples or $J_c$-ripples). This is not a mesoscopic, but a macroscopic phenomenon, which can be classified as a dissipative structure -- a result of the tendency for the spatially uniform physical systems driven away from thermal equilibrium to break the translation symmetry and form steady macroscopic patterns, e.g. sand ripples\cite{Hansen,Lan}, Taylor-Couette flow, thermal convection, oscillatory chemical reactions, etc.\cite{Cross}. The spatial scale of modulation is determined by the thermal diffusion length and in the case of coated conductors it is of the order of $0.3 -1\; cm$, much greater than the magnetic field penetration or coherence lengths.

\section{\label{sec:level1}Planar model\protect}
Coated conductors\cite{Larbalestier,Foltyn} are manufactured in the form of a tape in which the superconducting YBCO film of about $1\;\mu m$ thick is deposited on a buffered flexible metal substrate (e.g. $Ni-W$ alloy, Hastelloy or stainless steel). A copper stabilizer is either soldered or electroplated on top of the $YBCO$ film, Fig. 1. The typical width of such a tape-like wire is $4\;mm$, the thickness, about evenly divided between the substrate and stabilizer, is close to $100\;\mu m$ and continuous pieces with uniform critical current density are manufactured in lengths over $100\; m$ and up to $1\; km$..  

\begin{figure}
\includegraphics{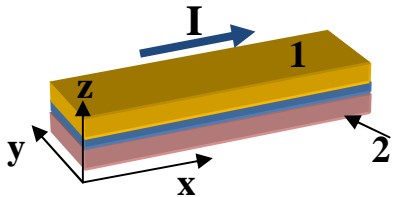}
\caption{\label{fig:} A sketch of the cross-section of coated conductor (not to scale). A thin superconducting film is sandwiched between the copper stabilizer (1) and metal substrate (2). The resistive interface between copper and $YBCO$ and the insulating buffer between $YBCO$ and substrate are not shown.}
\end{figure}

In recent literature the normal zone propagation in coated conductors has been treated as a three-dimensional (3D) problem with the help of finite element analysis software, e.g. Ref.\cite{Wan}. In this section we show that the properties of the coated conductors allow the reduction of this nominally 3D problem to a 2D or 1D problem without any significant loss of relevant physical content. Computationally, such reduction makes the problem much more tractable, especially when it comes to the analysis of the effects of the interfacial resistance that introduces strong non-linearity into the partial differential equations describing NZP.  

The reduction of 3D equations of heat conduction in a thin tape-like composite wire to a 2D (planar) or 1D (linear) model is possible if the variation of temperature across the thickness of the wire is negligible in comparison with its variation along the wire. The typical value of the current that coated conductor can carry is $J\sim 200-300\; A/cm\; width$. In the normal state the heat flux per unit area generated in the stabilizer 
\be
j_Q=\rho_1 J^2/d_1,
\ee
where  $\rho_1$ and $d_1$ are the resistivity and thickness of the copper stabilizer. At temperatures $77-90\;K$  $\rho_1\approx 0.2-0.3\;\mu\Omega\;cm$, and for $d_1=40\mu m$ we get $j_Q\sim 1\;W/cm^2$. In order to remove this heat through the surface of the conductor a temperature gradient across the thickness of the stabilizer is required:
\be
\f{K_1\Delta T}{d_1}\sim j_Q;\;\; \Delta T\sim \f{\rho_1 J^2}{K_1}
\ee 
The thermal conductivity of copper $K_1\sim 5 W/cm\;K$, so that the temperature variation across the thickness of stabilizer $\Delta T\sim 10^{-2}-10^{-3}\;K$. A similar  estimate can be made for the substrate. Due to its relatively high resistance, the amount of Joule heat generated in the substrate itself is negligible\cite{Levin}. The temperature gradient across the thickness of the substrate is necessary to transport the heat generated in the stabilizer. Thus, 
\be
 \f{K_2\Delta T}{d_2}\sim j_Q;\; \Delta T\sim \f{\rho_1 J^2d_2}{d_1K_1}.
\ee 
The thermal conductivity of Hastelloy $K_2\approx 7\times 10^{-2}W/cm\;K$ is substantially smaller than that of copper. Nevertheless, as long as the thermal flux $j_Q$ is of the order of $1 - 10\;W/cm^2$, the temperature variation across the substrate with thickness $d_2=50 - 100\;\mu m$ is still small, $\Delta T\sim 0.1-1\;K$, in comparison with the variation of temperature along the conductor during NZP. This conclusion remains true for two other types of substrate, $Ni-W$ alloy and stainless steel. 

The last structural element that needs to be examined in terms of its effect on thermal conduction is the buffer between the YBCO film and the substrate. The buffer has a total thickness $d_b\approx 150-200\; nm$ and consists of several layers of ceramic substances. If we take a representative value of the heat conductivity of Yttria stabilized Zirconia (YSZ) as $K_b\sim 1.5\times 10^{-2}W/cm K$, the temperature drop across the buffer that is required to transfer the flux $j_Q$ is
\be
\Delta T_b\sim \f{j_Qd_b}{K_b}.
\ee
Again, as long as the thermal flux $j_Q$ is of the order of $1 - 10\;W/cm^2$, the temperature drop $\Delta T_b\sim 10^{-3}-10^{-2}K$. The thickness of the interface between copper stabilizer and YBCO is even smaller and it is also can be considered thermally transparent. Thus, the temperature of the coated conductors can be considered as dependent only on in-plane coordinates $ T(\vec r)$, $\vec r=\{x,y\}$.

One of the specifics of coated conductors is that the superconducting film is much thinner than the stabilizer and the substrate. Therefore, it is natural to treat it as an infinitesimally thin layer, with negligible heat capacity capable, however, of carrying finite electric current density per unit width and generate a finite amount of heat per unit area. In a tape-like wire shown in Fig. 1 we place the superconducting film at $z=0$, while the stabilizer occupies the volume $0<z<d_1$, and the substrate is located at $-d_2<z<0$. Since the estimates presented above have shown that in coated conductors the redistribution of thermal energy and equilibration of temperature in the $z-$ direction is much faster than the heat propagation along the $\{x,y\}$ plane, the two-dimensional equation of heat conduction can be obtained straightforwardly from energy conservation. The energy balance in a cylindrical volume of height $d_1+d_2$ and arbitrary cross-section in the $\{x,y\}$ plane leads to the following equation:
\be
\f{\p U}{\p t}-\nabla\cdot \vec{j}_{Q} = -\left.j_{Q,z}\right |_{z=d_1} -\left.j_{Q,z}\right |_{z=-d_2}+Q.
\ee
Here $U$ is the internal energy of the conductor per unit of its surface area, and
\be
\f{\p U}{\p t}=\f{\p U}{\p T}\f{\p T}{\p t}\equiv C\f{\p T}{\p t},
\ee
where 
\be
C=C_1d_1 +C_2d_2
\ee
is the combined specific heat of the conductor. The contributions of the YBCO film, as well as that of buffer and the interface to the thermal mass of the conductor are negligible in comparison with that of the stabilizer and the substrate. 

The in-plane heat flux is defined as
\be
\vec{j}_{Q}=-K\nabla T,
\ee
where 
\be
K=K_1d_1+K_2d_2
\ee
is the effective in-plane thermal conductivity of the conductor. The heat flux from both, top and bottom surfaces, in the right-hand side of Eq. (5), we will take in the Fourier form
\bea
\left. j_{Q,z} \right |_{z=d_1} = K_0(T-T_0);\\\nonumber
\left. j_{Q,z} \right |_{z=-d_2} = K_0(T-T_0).
\eea
Here $K_0$ is the heat transfer coefficient across the insulation on the surface of the wire and $T_0$ is the ambient temperature which, in the absence of losses, is the operating temperature.  In the situations when the conductor is cooled with liquid coolant, the thermal flux from the surface has substantially more complex temperature dependence\cite{Ekin,G}. But here we are interested in the effect of the interfacial resistance, so it make sense to consider all other contributing factors in their simplest form. 
The areal density of the internal heat sources in Eq. (5)
\be
Q=\int_{-d_2}^{d_1}q(z)dz
\ee
is the sum of all internal heat sources. 
Thus, the 2D (in-plane) heat conduction equations for the coated conductor takes form
\be
C\f{\p T}{\p t}-\nabla\cdot (K\nabla T) =Q -2K_0(T-T_0).
\ee

The redistribution of current between the superconducting film and stabilizer is determined by the condition of charge conservation, 
\be
\nabla\cdot\vec{J_s}+ J_z=0;\; \nabla\cdot\vec{J_1}- J_z=0
\ee
where $\vec{J}_{s}$ and  $\vec{J}_{1}$ are the linear density of current[$A/cm$] flowing through the superconducting film and stabilizer respectively. The quench is a slow process in comparison to the time scale of charge redidtribution in a metal and the time derivative of the charge density in Eq. (13) can be neglected. The density of current flowing across the interface between the stabilizer and superconductor
\be
J_z=-\f{V_1-V_s}{\bar{R}},
\ee
where $V_1$ and $V_s$ are the local electric potentials of the stabilizer and superconductor, respectively and $\bar {R}\;[\Omega\; cm^2]$ is the resistance of the unit area of the interface. 

There are three internal heat sources - originating in the stabilizer, interface, and superconductor respectively:  
\bea
q=\f{1}{\rho_1}\vec {E}_1^2 +\f{(V_1-V_s)^2}{\bar {R}}\delta (z) +\vec{J}_s\cdot\vec{E}_s\delta (z).\nonumber
\eea
Here $\vec{E}_1=-\nabla V_1$ and $\vec{E}_s=-\nabla V_s$ are the electric fields in the stabilizer and superconductor, respectively. We use the delta-functions to account for the fact that two of the heat sources are concentrated in the volume much thinner than either the stabilizer or substrate. The substrate contributes only to the thermal mass of the conductor due to its large resistance\cite{Levin}.  
The integrated areal density of heat sources in Eq. (12) takes form
\be
Q=\f{d_1}{\rho_1}\vec {E}_1^2 +\f{(V_1-V_s)^2}{\bar {R}} 
+\vec{J}_s\cdot\vec{E}_s.
\ee

Equations (12-15), supplemented by the nonlinear constituent relationship between current and electric field in the superconductor are sufficient to solve the 2D NZP problem.
\subsection{\label{sec:level2} 1D problem}
NZP can be treated as a one-dimensional problem when the width of  the conductor is smaller or, at least, comparable to the thermal diffusion length, which we define below. In this case the thermal equilibrium along the width of the conductor (in the $y-$direction) establishes more quickly than along the conductor, so that one can only consider the evolution of $T(x,t)$, $J(x,t)$ and $J_s(x,t)$ along the conductor. Since $J_1+J_s=J=const$, the equations (13) reduce to one equation
\be
\f{\p J_s}{\p x}=\f{V_1-V_s}{\bar {R}}; 
\ee
which can also be used in the form
\be
\bar{R}\f{\p^2J_s}{\p x^2}= E_s-E_1.
\ee
The constituent relationship for a superconductor can be presented in many forms, the one most frequently used in literature is
\be
E_s(J_s)=E_0\left (\f{J_s}{J_c}\right )^n.
\ee
Here $J_c(T)$ is the critical current and $n$ is the exponent, usually large, $n\sim 20-40$. It is customary to take $E_0=1\mu V/cm$. For the stabilizer, the conventional Ohmic relationship will suffice:
\be
E_1(J_1)=\f{\rho_1}{d_1}J_1
\ee

The one-dimensional version of Eqs. (12), (15), together with Eq.(16) or (17), completely define the 1D NZP problem. However, computationally, these two coupled equations still present a fairly formidable problem. Below we will introduce an approximation that allows us to reduce the problem to one equation that still preserves the relevant physics of the phenomenon.
\subsection{\label{sec:level2} Reduction of 1D NZP problem to one equation}

\begin{figure}
\includegraphics{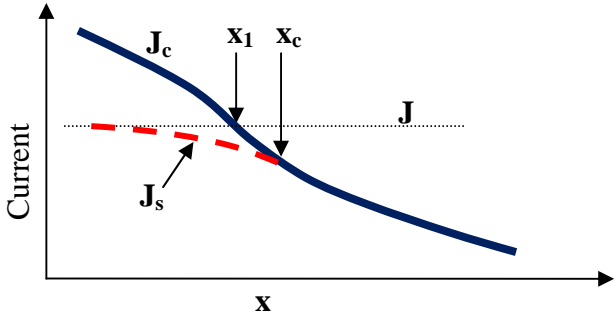}
\caption{\label{fig:} The solid line is a sketch of the temperature dependent critical current density $J_c$. Temperature increases from left to right. The dashed line $J_s$ is the current density that flows through the superconductor. The dotted line is the constant total current density. The point $x_c$ demarcates the boundary between the subcritical $E_s=0$ and critical $E_s>0$ sections.}
\end{figure}

In good quality $YBCO$ films the typical values of the exponent in Eq. (18) $n\approx 20—-40$ \cite{Ekin}. We will define the internal heat sources within the Bean model approximation\cite{Gurevich,Levin} which corresponds to the limit $n\rightarrow\infty $. Correspondingly, as long as the current density $J_s$ is less than the temperature dependent critical current density $J_c$, the electric field and dissipation in the superconductor is negligible. These we will call the subcritical sections of wire. In the critical sections of the conductor $J_s=J_c$. In the normal state, $T>T_c\;$, no current flows through the YBCO film. 

Let us consider the situation shown in Fig. 2. The temperature rises and, correspondingly, the critical current density $J_c(x)$ is falling from left to right. The point $x=x_1$ corresponds to the condition $J=J_c(x_1)$. Well to the left of this point, in the subcritical section, all current flows through the superconductor. However, in the vicinity of $x_1$ the current starts to leak out of the superconductor into the stabilizer. Therefore, the critical section defined by the condition $J_s=J_c(x_c)$ is located to the right of the point $x_1$ where $J=J_c(x_1)$. For $x<x_c$ (the subcritical section) $E_s=0$ and the condition of charge conservation, equivalent to Eq. (17) 
\be
\bar{R}\f{\p^2J_1}{\p x^2}= E_1- E_s;
\ee
takes form
\be
\bar{R}\f{\p^2J_1}{\p x^2}= \f{\rho_1}{d_1}J_1.
\ee 

The solution of this equation
\be
J_1=Ae^{\kappa (x-x_c)};\; \kappa \equiv 1/\lambda =(\rho_1/d_1\bar{R})^{1/2}
\ee
has two unknowns, $A$ and $x_c$. These can be found by matching the in-plane current 
\be
A=\left.J-J_c\right |_{x_c},
\ee
and out-of-plane current, see Eq.(13),
\be
\kappa A=\left . -\f{\p J_c}{\p x}\right |_{x_c}.
\ee
Let us introduce a dimensionless temperature $\theta$ and also assume for simplicity a linear dependence of the critical current on temperature\cite{Ekin,Gurevich}
\be
\theta=\f{T-T_1}{T_c-T_1};\;\;J_c(\theta )=J(1-\theta ).
\ee
The current sharing temperature $T_1$ is defined by the condition $J_c(T_1)=J$, so that $T(x_1)=T_1$ and, therefore, $\theta (x_1)=0$. Equations (23) and (24) take form
\be
A=J\left.\theta\right |_{x_c};\;\; \kappa A=J\left . \f{\p \theta}{\p x}\right |_{x_c}.
\ee
Thus, the current in the stabilizer in the subcritical section $x<x_c$ is given by 
\be 
J_1=J\theta(x_c) e^{\kappa (x-x_c)},
\ee
and in the critical section $x>x_c$
\be
J_1=J-J_c=J\theta .
\ee
The location of the critical point $x_c$ is determined by the condition 
\be
\kappa \theta =\pm \f{\p \theta}{\p x}.
\ee
Here the $\pm$ sign corresponds, respectively,  to either positive or negative derivative $\p\theta /\p x$. The solution of the Eq. (29), $\theta (x_c)\equiv \theta_c$, is always positive. Using the Taylor expansion $\theta (x_c)\approx \theta^{\prime}(x_1)(x_c-x_1)$ we get
\be
 x_c-x_1\approx \lambda .
\ee
Hereafter a prime indicates a spatial derivative.

Thus, in the subcritical region $ x<x_c$ the heat source, Eq. (15), is given by the sum of equal contributions from the stabilizer and interface 
\be
 Q=\f{\rho_1}{d_1}J_1^2 +\bar {R}(J_1^{\prime})^2 =\f{2\rho_1}{d_1}J^2(\theta(x_c))^2 e^{2\kappa (x-x_c)},
\ee

In the critical sections of the wire where $J_s=J_c$ all three internal heat sources appear - originating in the stabilizer, interface and superconductor respectively. From Eq.(16) follows the relationship between the electric fields in the stabilizer and superconductor
\be
J_c^{\prime }= (V_1-V_s )/\bar {R};\;\; E_s=E_1+\bar{R}J_c^{\prime\prime }
\ee
The heat source in the critical sections $x>x_c$ (Fig. 2) is given by
\bea
Q = \f{\rho_1}{d_1}(J-J_c)^2+\bar {R}(J_c^{\prime})^2 + \\ \nonumber 
J_c \left (\f{\rho_1}{d_1}(J-J_c) {}+ \bar{R}J_c^{\prime\prime}\right)=\f{\rho_1}{d_1}J(J-J_c) +\\ \nonumber \bar {R}(J_c^{\prime})^2 + \bar{R}J_cJ_c^{\prime\prime}.
\eea
Expressed in terms of the dimensionless temperature, Eq. (25), the heat source in the critical section has the form
\bea
Q =\f{\rho_1J^2}{d_1}\left [\theta +\lambda^2(\theta^{\prime})^2 - \lambda^2 (1-\theta )\theta^{\prime\prime}\right ].
\eea

Finally, in the normal sections where $T>T_c$ all current flows through the stabilizer and the heat source is given by
\be
Q= \f{\rho_1}{d_1}J^2.
\ee

In order to rewrite Eq.(12) in the standard dimensionless form and to reduce the large number of material constants to their relevant combinations, in addition to the definitions given by Eq.(25), we will measure the distance in units of thermal diffusion length $l_T =(D_T/\gamma)^{1/2}$, where $D_T = K/C$, and time in units of $\gamma^{-1}$, where the increment $\gamma =\rho_1 J^2/d_1C\Delta T$ determines the characteristic time required to warm an element of the conductor. Here $\Delta T \equiv T_c-T_1$

Then, piece-wise defined Eq. (12) at $T>T_c$ ($\theta >1$), together with Eq. (35), takes form
\bea
\f{\p \theta}{\p \tau}-\f{\p^2 \theta}{\p \xi^2} = 1-\kappa_0 (\theta -\theta_0);\;\;\theta >1;\\
\kappa_0=\f{2K_0\Delta Td_1}{\rho_1 J^2};\;\tau =\gamma t;\;\; \xi =x/l_T.\nonumber 
\eea

In the critical sections with temperature $\theta_c<\theta <1$ Eqs. (12) and (34) take the following form:
\bea
\f{\p \theta}{\p \tau}-\f{\p^2 \theta}{\p \xi^2} = \theta +r(\theta^{\prime})^2 - r(1-\theta )\theta^{\prime\prime}
-\kappa_0 (\theta -\theta_0);\\ \nonumber
\theta_c\leq\theta \leq 1; \;\;\;\;
\eea
Here $\theta_c\equiv \theta (x_c)$ is the floating boundary between the critical and subcritical sections determined by the condition (29).
The strength of the nonlinear terms is proportional to the interface resistance. 
\be
r=\f{\lambda^2}{l_T^2}=\f{\bar{R}}{R_0};\;\;R_0=\f{\rho_1 l_T^2}{d_1}=\f{K(T_c-T_1)}{J^2}.
\ee
Here $\lambda =(d_1\bar{R}/\rho_1)^{1/2}$ is the current transfer length, Eq. (22). 

Taking into account Eq. (31) we get for the subcritical regions $\theta<\theta_c$
\be
\f{\p \theta}{\p \tau}-\f{\p^2 \theta}{\p \xi^2} = 2\theta_c^2 e^{-2\kappa |x-x_c|}-\kappa_0 (\theta -\theta_0);\;\;\theta <\theta_c;
\ee

The equations (36)-(39), taken together, are piecewise defined dimensionless version of the Eq.(12). Although it is more complex than the standard Kardar-Parisi-Zhang (KPZ) equation\cite{Kardar}, they share a common feature -- the competition between diffusion and the nonlinear growth term $\propto (\theta^{\prime})^2$.  

Even more important is that Eq. (37) may have negative effective diffusion coefficient because it can be rewritten as follows:
\bea
\f{\p \theta}{\p \tau}-(1-r+r\theta)\f{\p^2 \theta}{\p \xi^2} = \theta +r(\theta^{\prime})^2 
-\kappa_0 (\theta -\theta_0);\\
 \theta_c\leq\theta \leq 1;\nonumber
\eea
For $r>1$ the effective nonlinear diffusion coefficient can be negative within the temperature interval $\theta_c<\theta<(r-1)/r$. Therefore, the necessary condition for a negative diffusion coefficient is 
\be 
r>\f{1}{1-\theta_c}>1,
\ee
where $\theta_c$ is determined self-consistently by Eq. (29). Thus, only when the interfacial resistance, Eq. (38), exceeds a certain threshold we can expect to see the pattern formation and other anomalous phenomena discussed below. 

To summarize, using the extreme limit of the constituent relationship (18), $n\rightarrow \infty$, we have reduced the system of the two equations, (12) and (13), to one piece-wise defined equation (36-39) for temperature. The price we have paid for that is that the boundary between the critical ($E_s>0$)  and subcritical ($E_s=0$)  regions $\theta_c$ is a floating one and has to be determined self consistently from Eq.(29). This presents a significant challenge for the numerical solution. Our next step is to simplify the problem by fixing the floating boundary at $\theta_c =0$. Comparing with the sketch in Fig. 2, this approximation means that for $x<x_1$ we take that all current flows through the superconductor, $J_s=J$, and for $x>x_1$ we take $J_s=J_c$.  This approximation becomes exact in the limit of small $\lambda \ll l_T$, see Eq. (30), and we assume that even for $\lambda \sim l_T$ the qualitative results will provide a meaningful guidance to the physics of the phenomenon. 

The equation (36) which describes the normal section of the conductor does not change. In Eq. (39) we set $\theta_c =0$. In Eq. (37) we need to ensure that all three heat sources defined by Eq. (15) are positive. Specifically it means that the electric field in the superconductor must have the same sign as the current. The direction of the current flow is taken to be positive. Therefore, $E_s=\rho_1J/d_1(\theta-\lambda^2\theta^{\prime\prime})$, see Eq. (32), must be positive or zero. Below we will use a step function to enforce this condition.  Equation (37) will be used in the form
\bea
\f{\p \theta}{\p \tau}-\f{\p^2 \theta}{\p \xi^2} = \theta^2 +r(\theta^{\prime})^2 +(1-\theta )(\theta - 
r\theta^{\prime\prime})H(\theta -r\theta^{\prime\prime})\nonumber \\
-\kappa_0 (\theta -\theta_0);\;\; 0\leq\theta \leq 1; \;\;\;\;
\eea
Here the unit step function $H(E_s)$ ensures that the electric field in the superconductor is either positive or zero. If the boundary between the critical and subcritical sections were determined exactly, this condition would be met automatically. But since we set the boundary fixed at $\theta_c=0$, the step function is necessary to prevent the numerical solution to "wander off" into unphysical territory where the electric field $E_s$ would have a direction opposite to current. In the range of temperature that meets the condition  $\theta -r\theta^{\prime\prime}\geq 0$ Eq. (42) takes the form of Eq. (37) or (40). In the sections where $H(E_s)=0$ the heat source is proportional to $\theta^2$ similar to that in Eq. (39). 
  
\section{\label{sec:level1}Results\protect}

\begin{figure}
\includegraphics{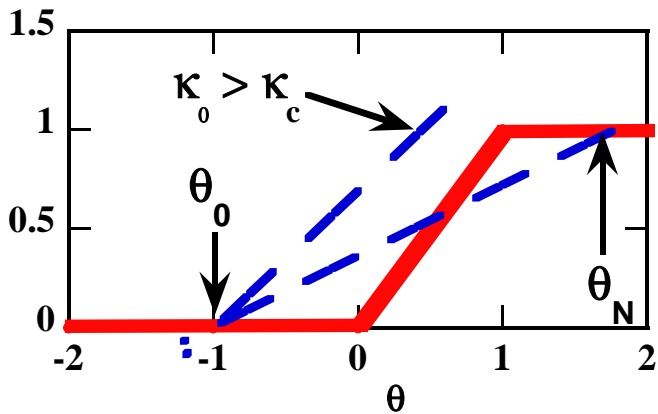}
\caption{\label{fig:} Sketch of the right-hand sides of Eqs. (36), (37), and (39) as a function of temperature in the limit of uniform temperature. Solid lines is the heat source, dashed lines indicate cooling term for two values of cooling constant $\kappa_0$.}
\end{figure}

The solutions of Eqs. (36), (39) and (42) with $\theta_c=0$ were obtained by two different methods in order to eliminate the possibility of computational artifacts.   We used a direct finite differences method and a high level software Mathematica.  The temperature profiles $\theta (\xi, \tau )$ presented below correspond to periodic boundary conditions and the initial condition in the form of a Gaussian in the center of the conductor 
\be
\theta (\xi, 0)=\theta_0 + (|\theta_0 |+a)\exp \{-\xi^2/2\delta^2\}. 
\ee
In all examples shown below we keep the width of the Gaussian on the order of the diffusion length, $\delta = \sqrt 2$.  
The dimensionless ambient (operating ) temperature is determined by the ratio of the transport current to the critical current at the operating temperature, Eq. (25),
\be
\theta_0=1-\f{J_c(T_0)}{J}<0.
\ee
The current sharing temperature is given by
\be
T_1=\f{T_c\theta_0-T_0}{\theta_0-1}.
\ee
The results shown below correspond to $\theta_0 =-1$ ($J=0.5J_c(T_0)$). Correspondingly, the current sharing temperature $T_1$ lies half-way between the operating and the critical temperature
\be
T_1=\f{T_c+T_0}{2}.
\ee
The maximum temperature of the initial temperature profile, see Eqs. (25) and (43), is given by
\be
T_{max}=T_c+(a-1)(T_c-T_1).
\ee
The results shown below correspond to the parameter $a=1.1$. The respective  maximum temperature of the initial temperature profile is slightly above $T_c$. The stability of the conductor with respect to various values of the maximum temperature was reported in Ref.\cite{ASC}.
The system described by Eqs.(36), (39) and (42) has two stable uniform ($\theta^{\prime }= \theta^{\prime \prime}=0$) modes of operation, see Fig. 3. One mode corresponds to the zero dissipation state of the conductor with temperature $\theta_0$. The other is a normal state with temperature
\be 
\theta_{N}=\theta_0 +\kappa_0^{-1}>1;\;\;T_{N}=T_1+(\theta_0 +\kappa_0^{-1})(T_c-T_1)
\ee
The condition of bistability is 
\be
\kappa_0 <\kappa_c =\f{1}{1+|\theta_0|}.
\ee
For stronger cooling, $ \kappa_0 >\kappa_c$, Eq.(36) does not have a stable uniform solution and there is only one stable uniform mode of operation -- the superconducting state with temperature $\theta_0$. 
\begin{figure*}
\includegraphics{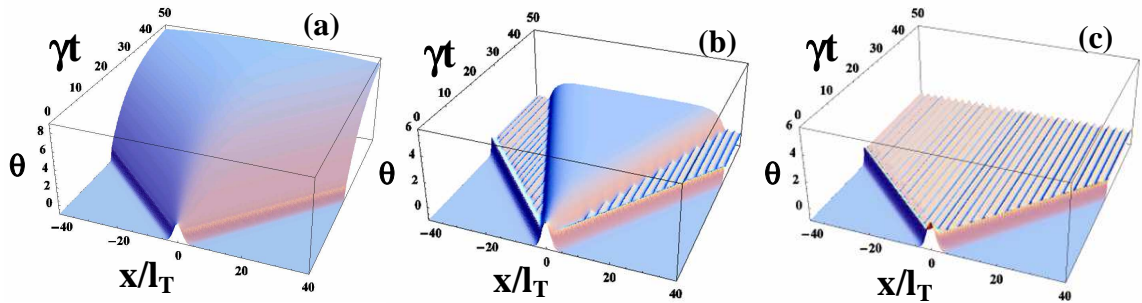}
\caption{\label{fig:wide} Sequence of temperature profiles as they develop from the initial Gaussian perturbation visible in the foreground. All profiles correspond to the interface resistance $r=1.4$, Eq. (38). The width of the initial temperature profile $\delta =1.4$, Eq. (43). The range of the plots are $0<\gamma t \leq 50$ and  $-40\leq x/l_T\leq 40$. (a) $\kappa_0 =0.1$; Conventional normal zone, but it propagates with greater speed than that at $r\ll 1$. (b) $\kappa_0 =0.35$; Conventional normal zone and temperature ripples emerge simultaneously, but propagate with different speed. (c) Cryostable condition ($\kappa_0 =0.55>\kappa_c$). Only the temperature ripples are triggered.}
\end{figure*}

The solutions of Eqs.(36), (39) and (42) with $\theta_c=0$ fall into several categories (scenarios) determined by the three dimensionless parameters $\{r,\kappa_0, \theta_0\}$. Here we present a few scenarios that are important from both the basic physics and applications point of view. In all cases discussed below we take the transport current $J=0.5J_c(T_0)$, which corresponds to $\theta_0 =-1$ and $\kappa_c =0.5$. Figure 4 shows the 3D plots of the spatiotemporal development of the initial Gaussian temperature profile for different values of the cooling constant $\kappa_0$ and the interface resistance above the threshold $r=1$ (negative effective diffusivity, Eq. (41)). 

In Fig. 4(a) $\kappa_0 = 0.1$ (weak cooling, $\theta_{N}=9)$. This solution of Eqs. (36,39) and (42) describes a conventional bi-stable operation of the stabilized superconducting wire.  Additional heat generated in the interface (KPZ-type growth) substantially increases the speed of normal zone propagation. The detailed analysis of the effect of the increased interfacial resistance on propagation speed and wire stability is described in Ref.\cite{ASC}.  

At stronger cooling, but still in the bistable regime ($\kappa_0 = 0.35<\kappa_c$), an anomalous third mode of conductor operation emerges as shown in Fig. 4(b). The initial Gaussian profile gives rise to two distinct fronts propagating with a different speed. A more rapidly propagating front leaves in its wake a metastable state characterized by the static spatial temperature modulation (T-ripples) with the peak temperature below $T_c$. The temperature modulation causes a modulation of the critical current density, which is why we may also call this pattern $J_c$-ripples. The $J_c$-ripples in turn are absorbed into the conventional normal state slowly propagating on top of them. 

When $\kappa_0 \geq \kappa_c$, see Eq.(49) and Fig. 3, the cooling power at all temperatures is greater than the maximum power that can be dissipated in the stabilizer and, as the result,  the uniform normal state is unstable. However, as Fig. 4(c) demonstrates, for $r>1$ such cryostable, by conventional criterion, conductor is in fact a bi-stable system with the $J_c$-ripples being the second steady mode of operation. The initial temperature profile evolves into the $J_c$- ripples so that the  average temperature of the conductor behind the front remains steady, close to $T_1<T_c$, instead of relaxing to the uniform superconducting state with temperature $T_0$. 

\subsection{\label{sec:level2} Properties of $J_c$-ripples}

In Fig. 5 the temperature profile of the $J_c$- ripples mode is shown. At a distance of a few diffusion lengths behind the propagating front the temperature modulation is stable and alternates in the range $-0.8 < \theta < 0.26$. Inspite of the overall symmetry of the problem, small computing errors and interpolation procedures lead to difference in the propagation speed of the two fronts and some irregularities in the profile. On the smaller scale, however, the profile is rather regular as shown in the inset. The length of modulation is $2.7 l_T$. 

A characteristic feature of these ripples is an apparent discontinuity of the first derivative $\p\theta /\p x$. This allows us to determine the value of the peak temperature $\theta_p\equiv \max \{\theta (x)\} $. Integrating Eq. (40) over an infinitesimal interval that includes the temperature peak we get 
\be
(1-r+r\theta_p)\left |\f{\p\theta}{\p\xi}\right |_{peak}=0.
\ee
Since the derivative is finite, this condition can be satisfied by 
\be
\theta_p=(r-1)/r
\ee
This corresponds to zero value of the effective thermal diffusivity in Eq. (40). In numerical calculations the second derivative remains finite and the value of $\theta_p$ remains slightly below the limit given by (51).
\begin{figure}
\includegraphics{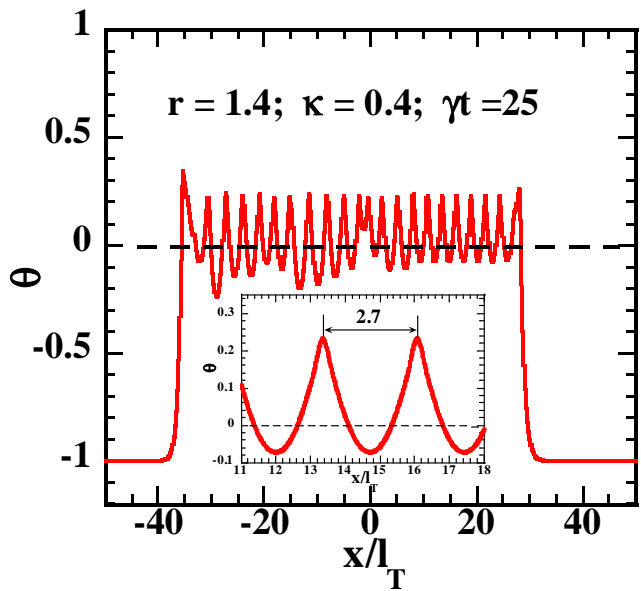}
\caption{\label{fig:} Profile of the $J_c$-ripples ($r=1.4$, $\kappa_0=0.4$) at a certain moment of time, $\gamma t=25$. The inset shows a magnified view of the same profile. The periodicity of the  $J_c$-ripple is $\approx 2.7\;\l_T$. The peak temperature is $\theta_p\approx 0.24-0.26$ in agreement with the exact value $1-r^{-1}$, Eq. (51).}
\end{figure}

The heat generating regions ($\theta >0$) are spaced along the wire, segregated by the colder regions sustained by heat diffusion. One can compare the average power dissipation in the $J_c$-ripples mode and in the conventional mode. A section of the conductor between two minima of the temperature is thermally insulated from the rest of the system because the thermal flux vanishes at these points. Integrating Eq. (12) over the distance $\Delta L$ between two troughs we get
\be
\f{1}{\Delta L}\int Qdx =2K_0 (T_{av}-T_0);\;T_{av}\equiv \f{1}{\Delta L }\int T(x)dx
\ee
On the other hand, the power dissipation in the normal mode
\be
Q_N=\f{\rho_1J^2}{d_1}=2K_0(T_N-T_0)  
\ee
Since $T_N>T_{av}$ (see Eq. (48)), the average power dissipated by the $J_c$-ripples mode is smaller than that dissipated in the normal mode. In other words, in the regime where current is repeatedly exchanged between the superconductor and stabilizer the conductor behaves, at least in terms of losses, as a hyperconductor –- a hypothetical substance with the resistance that is finite, but lower than that of the stabilizer, including the one made of copper.

\begin{figure}
\includegraphics{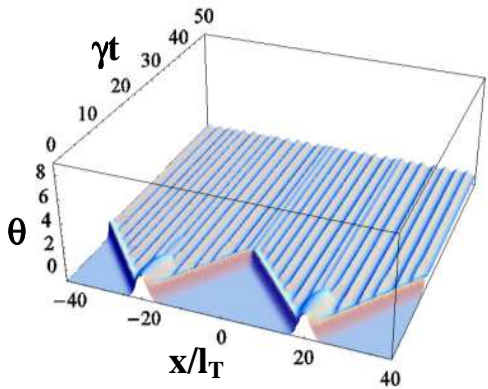}
\caption{\label{fig:}The initial condition gives rise to two propagating fronts that eventually merge. Here $r=1.5$, $\kappa_0=0.4$, and $\theta_0=-1$. There is a domain boundary between the two merged fronts with slightly irregular temperature variation.}
\end{figure}

Figure (6) illustrates how two spreading domains of ripples, originating from different sources, interact. The initial condition is two Gaussians which do not overlap. Each of them gives rise to the spreading $T$-ripples. Once the spreading domains merge, a stable domain boundary characterized by a slightly irregular temperature variation is formed.

\subsection{\label{sec:level2} External heat source and stability}

In order to study the stability of the ripples pattern we added a  ``heater''  -- an external heat source in the right-hand sides of Eqs. (36),(39) and (42) that can be switched on and off at the moments $t_1$ and $t_2$ respectively:
\be 
Q_{ext}=b\exp\{-\xi^2/2\delta_1^2\}H(t-t_1)H(t_2-t).
\ee
Such a source simulates a typical experiment on normal zone propagation in which a small resistive heater is thermally anchored to the superconducting wire in order to trigger a transition to the normal state\cite{Wang}. 

Figure (7) illustrates the following scenario. Unlike in Fig. 4(b), the transition to the ripples mode is triggered by the initial condition without tripping the conductor into the normal state. Since the cooling constant $\kappa_0 =0.4$ is below the cryostability criterion ($\kappa_0 <\kappa_c=0.5$),  the stable normal state with $\theta_{N}= 1.5$ is still possible. This indicates that there is a finite separation between the ranges of stability of the ripples mode and normal mode -– one can be triggered without triggering another. At the moment $\gamma t=20$ the heater is turned on and turned off at $\gamma t=25$. A characteristic ``rhino horn'' of rising temperature triggers the secondary transition to the normal state. Thus, in this range of parameters, the conductor is a tri-stable, rather than a bi-stable system. 

\begin{figure}
\includegraphics{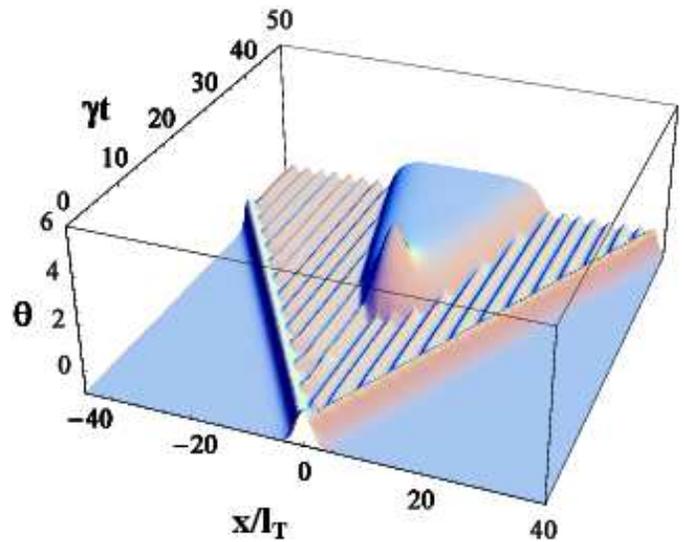}
\caption{\label{fig:} Tri-stable operation which  corresponds to  $\kappa_0 =0.4$,  $r=1.4$, and $\theta_0=-1$. The initial profile evolves into $J_c$-ripples. A heater, Eq. (54), is switched on at $\gamma t=20$ and switched off at $\gamma t=25$. It triggers the transition to the normal state. The evolution of the system depicted in the Figure corresponds to $b=1.5$, $\delta_1 =1.4$
}
\end{figure}

An ability to trigger and observe the temperature ripples in an experiment depends, among other things, on stability and robustness of this pattern. A scenario shown in Fig. 7 has demonstrated a finite margin of stability of the ripples with respect to transition to the normal state. In Fig. 8 the effect of a `` cold finger ''  -- a negative power source is shown. Physically, this is equivalent to bringing the conductor in contact with a thermal mass with a temperature lower than $T_0$. The cold finger was modeled by a Gaussian, similar to that given by Eq. (54), but offset from the center and with a negative value of the parameter $b$. In Fig 8 $b=-3$, $\delta_1 =8$, $\gamma t_1 =20$, and $\gamma t_2 =35$. After the application of the cold finger the temperature in its vicinity collapses well below the background temperature $\theta_0 =-1$. The cold finger catches and holds the nearby propagating front, preventing it from expanding. Once the cold finger is lifted, the front resumes its propagation with the same speed as before. This demonstrates that the  $T$-ripples mode is robust and recovers after severe disruptions.  

\begin{figure}
\includegraphics{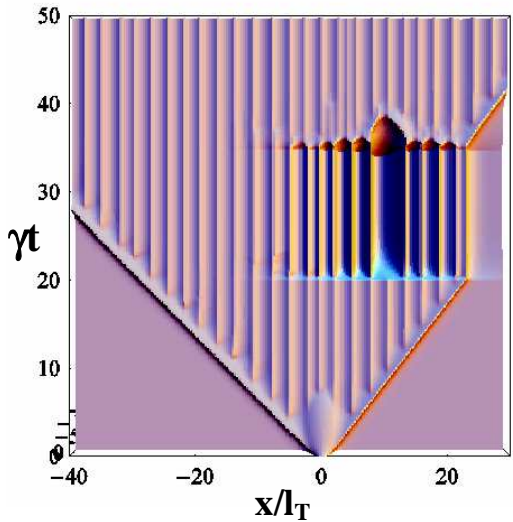}
\caption{\label{fig:}Effect of cold finger. Shown here is view ``from above''.  The transition to $J_c-$ripples mode is triggered by the initial Gaussian profile. The negative power source collapses temperature in its vicinity, but does not destroy the ripple mode. Parameters here are the same as in Fig. 6. 
}
\end{figure}

\subsection{\label{sec:level2} Stationary solutions}
\begin{figure}
\includegraphics{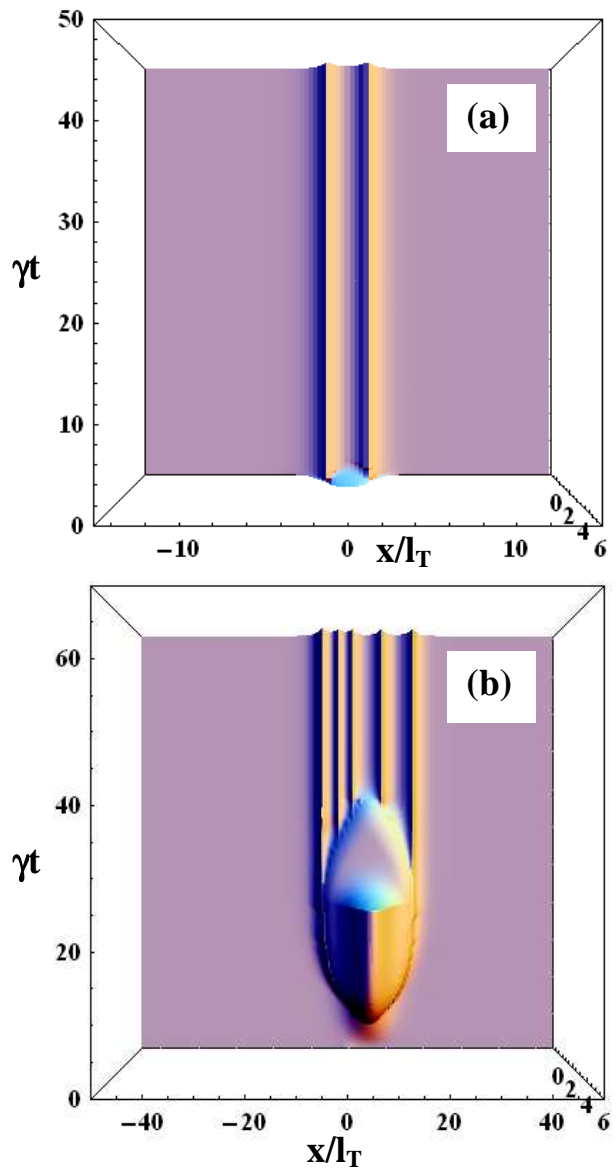}
\caption{\label{fig:} View ``from above''.  (a) A stationary bimodal solution that evolves from the initial Gaussian perturbation. (b) A more complex profile emerging under the action of the heater, Eq. (54). In both cases $r=1.2$, $\kappa_0=0.5$, and $\theta_0=-1$.   
}
\end{figure}

The previous scenarios depicted in Figs. (4-8) show propagating fronts. Equations (36), (39) and (42) also have stationary solutions. A stable temperature profile can be established either from an initial perturbation or by the action of an external heater. Fig. 9(a) shows a simple bimodal structure similar to that shown in the inset to Fig. (5). 
This bimodal soliton evolves from the initial Gaussian profile. A more complex five-peak structure shown in Fig. 9(b) evolves under the action of the external heater, Eq. (54), with the same width as the initial condition in Fig. 9(a). In both cases the control parameter $r=1.2$, Eq. (38), is the same, as are the cooling constant $\kappa_0=0.5$ and the ambient temperature $\theta_0=-1$, Eq. (44). 
\begin{figure}
\includegraphics{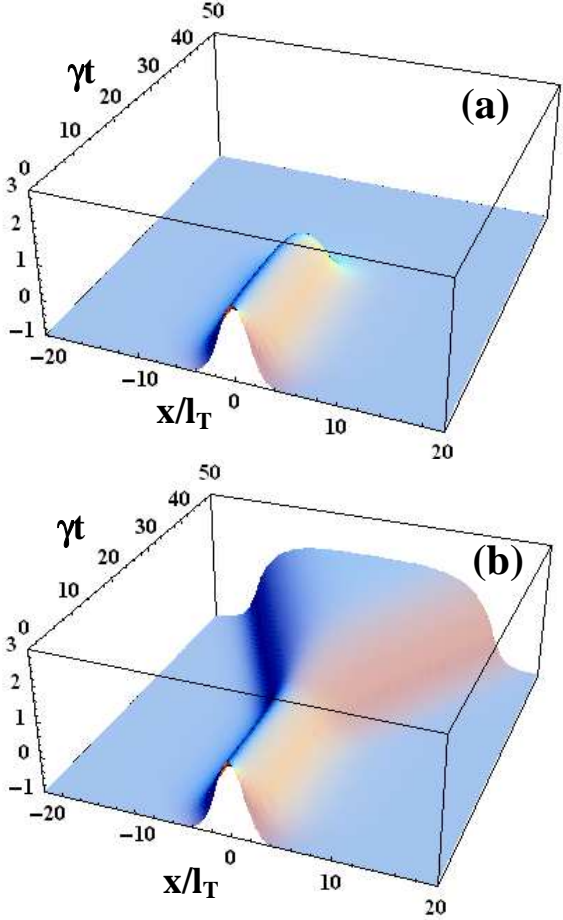}
\caption{\label{fig:}(a,b) 
Shown are long lived perturbations evolving from the initial Gaussian profile with $\delta =1.4$. (a) The value of $r=1.0426$ is very close to threshold $r=1$ above which the effective diffusivity can be negative. The profile lingers for a long time, but eventually disappears. (b) At slightly greater $r=1.0427$ the normal zone emerges and starts to propagate. In both cases $\kappa_0 =0.35$, $\theta_0=-1$ and $\delta =1.4$. 
}
\end{figure}

The stable solitons shown in Figs. 9(a,b) can be contrasted with metastable –- long lived perturbations shown in Fig. 10. In this scenarios the control parameter $r$ is very close to the threshold $r=1$. In Fig. 10(a) $r=1.0426$. The perturbation evolving from the initial Gaussian profile is quasistable and persists for about $\gamma t\approx 22$ before dissipating. A very small increase of the interfacial resistance to $r=1.0427$ yields a long lived perturbation that eventually gives rise to normal zone propagation. This is consistent with our previous observation that an increase of the interfacial resistance reduces cryostability of the current carrying conductor: thermal perturbations that dissipate at lower values of resistance give rise to a NZP at higher resistance\cite{ASC}.
What is important to realize comparing Figs. (9) and (10) is that it is the nonlinear KPZ term $\sim (\theta^{\prime})^2$ and the negative effective diffusivity in Eq. (37) that allow the isolated solitons like the ones in Fig. (9) to be stable and robust solutions over a finite range of parameters.

\subsection{\label{sec:level2} Thermal oscillations}
Physical systems described by the parabolic partial differential equations do not support the propagations of waves. It is instructive, however, to investigate whether the nonlinearities may change that and allow a long range propagation of time-dependent perturbations originating from a localized source.  The external power source, Eq. (54) was modified as follows
\be 
Q_{ext}=b\exp\{-\xi^2/2\delta_1^2\}(1+\sin (2\omega t))(1-e^{-t/\tau_0}).
\ee
This is equivalent to a resistive heater supplied with AC current with frequency $\omega$ and switched on gradually over a period of time $\sim \tau_0$. The results are presented in Fig. 11. 
\begin{figure}
\includegraphics{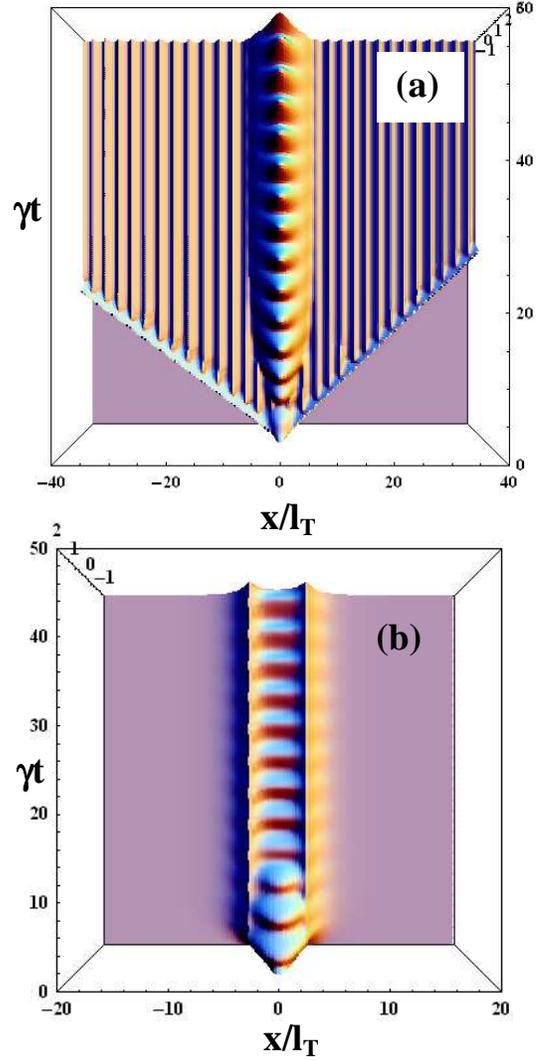}
\caption{\label{fig:} (a) The ripple mode is triggered by the initial Gaussian profile. The pulsations of temperature with frequency $2\omega = 1.5 \gamma $ are caused by the heat source given by Eq.(55). Here $r=1.5$, $\kappa_0=0.5$, $\theta_0=-1$, $\delta =\delta_1=1.4$, $b=1$. (b) The initial Gaussian profile gives rise to a bimodal stationary soliton. The temperature pulsations are confined within the walls of the soliton. Here $r=1.2$, $\kappa_0=0.55$, $\theta_0=-1$, $\delta =\delta_1=1.4$, $b=0.3$.
}
\end{figure}

In Fig. 11(a) the ripple mode is triggered by an initial temperature profile and the pulsating power source is gradually switched on with the time constant $\tau_0=10\gamma^{-1}$. It is evident that the pulsations do not propagate outside the range of localization of the source. If there were a phenomenon of `` thermal sound '' it would manifest itself as an undulation of the peak temperature lines that run along the time line. However, these lines are perfectly straight. 

Figure 11 (b) illustrates how the confinement of the thermal perturbations takes place. The control parameter $r=1.2$ and the cooling constant $\kappa_0=0.55$ are chosen to yield a bimodal stationary soliton similar to that in Fig. 9(a). The variable external power source creates temperature pulsations that remain strictly confined between the temperature peaks. This indicates that the singularities caused by the negative effective diffusivity block the propagation of small perturbations through them.

\begin{figure}
\includegraphics{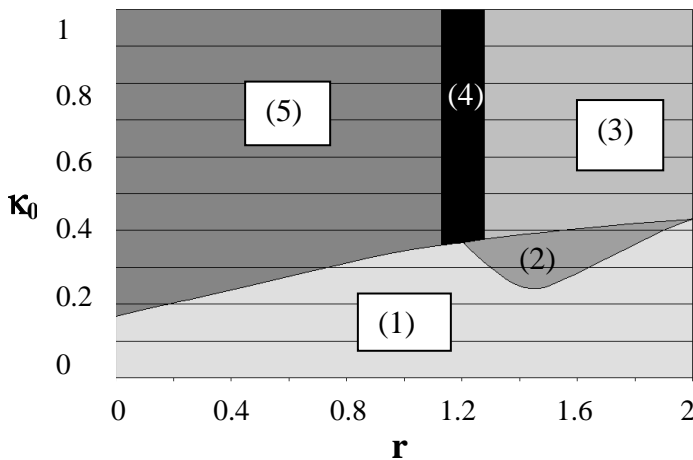}
\caption{\label{fig:} The range of parameters $\{r,\kappa_0\}$ specific for different regimes. The value of $\theta_0=-1$ and the width of the initial perturbation $\delta =1.4$ are the same for all combinations of $\{r,\kappa_0\}$. Area (1) corresponds to conventional NZP. In area (2) $T-$ripples and NZP coexist. In area (3) $T-$ripples are the second mode of operation. Area (4) corresponds to stationary solitons, and area (5) is the range of cryostability. 
}
\end{figure}

\subsection{\label{sec:level2} Phase diagram}

Figure 12 presents an approximate phase diagram that indicates the range of parameters $\{r,\kappa_0\}$ where different regimes take place. This diagram corresponds to a particular value of $\theta_0=-1$ which corresponds to transport current equal $50\%$ of the critical current at ambient temperature $T_0$, see Eq. (44). It needs to be mentioned that the boundaries between the regimes depend not only on three parameters $
\{r,\kappa_0,\theta_0\}$,  but on the width of the initial perturbation as well. This is obvious because the heat generated in the interface is determined by the derivative of temperature (the KPZ term in Eq. (42)). The  diagram shown in Fig. 12 corresponds to the width of the Gaussian profile, Eq. (43), $\delta =1.4$. 

In the range of parameters indicated as area $\#(1)$ in Fig. 12 the transition from superconducting mode to normal takes the form shown in Fig. 4(a) –- a conventional normal zone propagation, albeit with increased propagation speed and reduced stability margins in comparison with that at low interfacial resistance. 

Area $\#(2)$ corresponds to simultaneous emergence of two propagating fronts as shown in Fig. 4(b). Area $\#(3)$ is a range of relatively strong cooling and interfacial resistance above the threshold $r=1$. In this range of parameters the conductor is bistable with $T-$ripples being the second stable mode of operation, Fig. 4(c). A strip $\#(4)$ corresponds to the stationary solitons, Fig. 9. Area $\#(5)$ is the range of cryostability –- the temperature perturbations dissipate without triggering a transition to the normal state. Figures 10(a,b) illustrate the evolution of the initial temperature profile on both sides of the border between areas $\#(1)$ and $\#(5)$.

\section{\label{sec:level1} Experimental conditions\protect}

In order to estimate the interface resistance required for the emergence of $J_c$- ripples we take the values of physical parameters from Ref.\cite{Ekin}. Let us take the operating temperature $T_0 = 67\;K$ and $T_c\approx 87\;K$. At $J=(1/2)J_c^{(0)}$, $T_1=(T_0+T_c)/2=77\; K$. The specific heat of copper and substrate at $77\;K$ are, respectively,  $C_1 \approx 1.7\;J/cm^3\;K$ and $C_2\approx 1.4\;J/cm^3\;K$. Let us take $d_1=40\mu m$ and $d_2 =50\mu m$. Then, $C=C_1d_1 +C_2d_2\approx 14.2\times 10^{-3} J/cm^2\;K$. The thermal conductivity of copper and substrate at $77\;K$ are, respectively,  $K_1\approx 5W/cm\;K$ and  $K_2\approx 7\times 10^{-2} W/cm\;K$. Thus, $K=K_1d_1+K_2d_2\approx 20.3\times 10^{-3}W/K$. The thermal diffusivity of the coated conductor $D_T=K/C\approx 1.4\;cm^2/s$. 

The self-field critical current density of the state of the art coated conductors at $T=67\;K$ can be close to $J_c^{(0)}\approx 400 \;A/cm$, and, correspondingly, the transport current density at $50\%$ capacity can be $J=200\;A/cm$. The sheet resistance of the stabilizer is $\rho_1 /d_1\approx 0.5\times 10^{-4}\Omega$.  Taking $\Delta T =10\;K$, we obtain the increment $\gamma =\rho_1 J^2/d_1C\Delta T \approx 14\;s^{-1}$. The thermal diffusion length $\l_T=(D_T/\gamma )^{1/2}\approx 3\;mm$. The characteristic speed of NZP can be estimated as
\be
U_T=l_T\gamma\approx 4\;cm/s.
\ee
When the interfacial resistance is negligible, $\lambda\ll l_T$, the NZP speed is close to $u_T$.  This relatively slow propagation speed complicates quench detection and quench protection in coils made out of coated conductors. Increasing interfacial resistance leads to increasing NZP speed and reduced stability\cite{ASC}.
 
In order to observe the $J_c$- ripples the interface resistance has to exceed the characteristic value (Eq. 38)) 
\be
R_0=\f{\rho_1 l_T^2}{d_1}\approx 5\times 10^{-6}\;\Omega cm^2. 
\ee
The interface resistance of the currently manufactured coated conductors is about $50\;n\Omega cm^2$\cite{Polak}. Thus, two orders of magnitude increase in interface resistance is needed in order to create conditions under which the $J_c$- ripples may be observed. This can be readily accomplished by modifying the stabilizer application procedures, e.g. Ref.\cite{Duck}.

Taking into account the definition of the increment $\gamma$, the cooling constant $\kappa_0$ defined by Eq. (36) can be rewritten as
\be
\kappa_0=\f{2K_0\Delta Td_1}{\rho_1 J^2}=\f{2K_0}{C\gamma}.
\ee
To attain the value of $\kappa_0\approx 0.5$, see Fig. 12, one needs to have the heat transfer coefficient
\be
K_0\approx \f{C\gamma\kappa_0}{2}\approx  5\times 10^{-2}\;W/cm^2\;K.
\ee
For a typical insulator like Kapton\cite{Ekin} the thermal conductivity at $77\;K$ 
\be
h\approx 1.3\times 10^{-3}\;W/cm\;K.
\ee 
Since 
\be
K_0=\f{h}{d_0},
\ee
where $d_0$ is the thickness of insulation on the surface of the conductor, it is sufficient to have $d_0\approx 200 - 300\;\mu m$ to achieve the desired range of $\kappa_0$ where one can expect to see the anomalous phenomena described above.
\section{\label{sec:level1} Summary\protect}

In summary, we have uncovered a novel pattern of superconducting wire operation in which the DC transport current is shared between the superconductor and stabilizer. This phenomenon manifests itself through spontaneously developing temperature and critical current modulation along the wire with the spatial scale of the order of $1 \;cm$. Although the physics of this phenomenon is very different, the appearance of the temperature ripples is remarkably similar to that of the sand ripples. Most notably, this similarity extends to an apparent discontinuity of the slopes at the peaks\cite{Hansen,Lan}. 

\section{\label{sec:level1} Acknowledgments \protect}

J.P. R.  was supported in part by the Air Force Office of Scientific Research under grant no. FA9550-06-1-0479.
%\bibliography{apssamp}

\end{document}